 \documentclass[preprint2]{aastex}

\slugcomment{Not to appear in Nonlearned J., 45.}

\shorttitle{The ELODIE archive}
\shortauthors{Moultaka et al.}

\begin{document}


\title{The ELODIE archive\thanks{http://atlas.obs-hp.fr/elodie}}


\author{J. Moultaka\altaffilmark{}}
\affil{Physikalisches Institut, Z\"ulpicher Str. 77, D-50937 K\"oln, Germany\\
\&  LUTH, Observatoire de Meudon,
           5, place Jules Janssen,
           F-92190 Meudon cedex, France}
\email{moultaka@ph1.uni-koeln.de}

\author{S. A. Ilovaisky\altaffilmark{}}
\affil{Observatoire de Haute-Provence (CNRS), F-04870 Saint-Michel l'Observatoire, France}
\email{ilovaisky@obs-hp.fr}

\author{P. Prugniel\altaffilmark{}}
\affil{CRAL, Observatoire de Lyon,
            9, av. Charles Andre,  F-69561 St.Genis Laval cedex, France\\
         \&
          GEPI, Observatoire de Meudon,
           5, place Jules Janssen,
           F-92190 Meudon cedex, France}
\email{prugniel@obs.univ-lyon1.fr}
\and

\author{C. Soubiran\altaffilmark{}}
\affil{Observatoire Aquitain des Sciences de l'Univers, 
           UMR 5804, BP 89, F-33270 Floirac, France}
\email{Caroline.Soubiran@obs.u-bordeaux1.fr}




\begin{abstract}
The ELODIE archive contains the complete collection of high-resolution echelle spectra accumulated over the last decade using the ELODIE spectrograph at the Observatoire de Haute-Provence 1.93-m telescope. This article presents the different data products and the facilities available on the web to re-process these data on-the-fly. Users can retrieve the data in FITS format from http://atlas.obs-hp.fr/elodie and apply to them different functions: wavelength resampling and flux calibration in particular. 
\end{abstract}





\section{Introduction}\label{sec:intro}
Archiving astronomical data has always been a traditional activity of ground-based observatories. However, in the last 20 years this traditional responsibility of data producers has been significantly weakened. While photographic plates have been carefully stored and archived in the past, magnetic tapes have not always been kept for more than a few years and much data have thus disappeared. Photographic plates, and in particular spectra, are often still safely kept and efforts to make them publicly available are being deployed (Griffin 2004), but serious archiving of electronic data was only developed in the wake of the HST archive.

Now, with the upcoming Virtual Observatory and multi-wavelength,
multi-epoch or multi-parametric analysis, there is renewed interest in this material and consequently there is a large demand for on-line availability of  observational data. Archiving is also a way to optimise observing time (it avoids redundancy in observing programs) and ensures a better scientific efficiency for an instrument.

In this paper we present the on-line archive of the data obtained with the ELODIE spectrograph which has been in operation at the Observatoire de Haute-Provence (OHP) 1.93-m telescope since mid-1993. 
The main characteristics of the instrument are described in Baranne et al. (1996) and some of them listed in Table~\ref{tab:car}. The on-line archive is available since September 2003 and allows access to the FITS data and to the pipeline which re-processes them on-the-fly. 

In Sect. 2 we present the content of the archive. The original archived data are described in Sect.~3 and the organisation of the archive in Sect.~4. Section 5 describes the user interface, the data processing and the different data products. Plans for future development are discussed in Sect. 6.





\section{Statistics and scientific content}
The archive presently contains more than 18000 spectra obtained in the interval 1994 - 2001. The data which are accessible through the archive  have passed the period of exclusivity (of two years). This period is imposed by OHP for all observers except those who requested a longer period of protection for their data. After this period, data become available to the astronomical community.\\
The archived spectra are only those which have a signal-to-noise ratio (S/N) per pixel at 5500\AA\ larger than 10. 
The characteristics of the content are shown in Table ~\ref{tab:car}. 


The scientific content of the archive is very rich, since different kinds of objects have been observed with ELODIE for different purposes. The high accuracy of the radial velocities obtained with this instrument (now reaching $\sim 7m s^{-1}$ in the scrambled-fiber, simultaneous-Thorium mode) allowed the discovery of the first exoplanet around 51~Peg by Mayor \& Queloz (1995). The search for exoplanets is continuing with ELODIE. The discovery of spectroscopic and eclipsing binaries as well as M-dwarf systems and the determination of stellar masses (Delfosse et al. 1999) has also been made with ELODIE thanks to the accurate radial velocities.

The spectrograph has been used as well for asteroseismology (Bertaux et al. 2003 etc) and to determine stellar abundances and stellar parameters (Erspamer \& North 2002,2003, Kovtyukh et al. 2003, Soubiran et al. 2003, Santos et al. (2003), Katz et al. 1998, Oblak et al. 2002, Monier et al. 2002 etc...). 
Research using the off-line archive  allowed the development of software for automatic determination of atmospheric stellar parameters 
Teff, log(g) and [Fe/H] (TGMET, Katz et al. 1998). 

The compilation of the reference library for the TGMET software by Soubiran et al. (1998) was the first step to make reduced ELODIE spectra available through the web. This opportunity was used for instance by Mishenina \& Kovtyukh (2001) who selected in the library a large set of metal-poor stars to measure abundances of various elements.

In addition, different studies of open clusters (Soubiran et al. 2000), planetary nebulae (Neiner et al. 2000), RR Lyrae (Chadid et al. 1999), Herbig Ae/Be stars (Corporon \&  Lagrange-Henri 1999), comets (The Hale-Bopp comet by Rauer et al. 1997), long period variable stars (Mira variables) (ex. Alvarez et al. 2000) and X-ray source counterparts  (Chevalier \& Ilovaisky 1997) have also been made with ELODIE. 

Moreover, flux calibrated stellar libraries have been created (Prugniel \& Soubiran 2001, Imbert 2002) and used for stellar population synthesis at high spectral resolution (Le Borgne et al. 2002). 

Finally, observations with ELODIE in the Milky Way allowed to determine the distribution of stars in the Galactic Disk (Siebert et al. 2003) and the distribution of widths and intensities of the Diffuse Interstellar bands (DIB) (Tuairisg et al. 2000).

\section{The original archived data}\label{sec:archiveddata}

The unique characteristic of both ELODIE and its twin spectrograph  CORALIE at ESO (La Silla) in Chile is that the special cross-disperser combining a prism and a grism yields near constant inter-order spacing, allowing 67 orders to fit into a compact CCD 1024x1024 data format.  

Two entrance apertures (each 2 arcsec wide) in the focal plane of the telescope are dedicated respectively to the simultaneous observation of the target and of either the sky or the thorium lamp which is used to derive the wavelength calibration. The continuum lamp (a tungsten lamp) frames are used to derive the location of the 67 orders and to extract both the spectrum of the flat field and the blaze function. 

The original archived data include the results of a reduction process performed automatically at the telescope, immediately after  data acquisition. This reduction includes correction for bias, dark or scattered light subtractions, cosmic-ray rejection and optimal extraction of the data along the orders (using Horne's algorithm). It includes as well the division by the spectrum of the flat-field and by the blaze function (both normalised to unity in each order), followed by wavelength calibration. The results of this original processing are not entirely satisfactory (as explained below) and consecutive echelle orders could not be reconnected.

\section{Organisation of the arhive}\label{sec:organi}

The ELODIE archive is organised in three main structures:
\begin{itemize} 
\item The first one contains the original data copied onto a dedicated machine and itemized in a PostgreSQL (http://www.postgresql.org/) database managed by the SQL language.
\item The second one holds the code package written in C or Fortran 77 proper to the ELODIE archive. These programs also use routines from the PLEINPOT software package  distributed by the MIGALE project\footnote{http://www.cai-mama.obspm.fr/migale/va/englishGroupeProjet.html} and developed by Ph. Prugniel and the MIGALE group.
\item The third structure is the web interface written in HTML and Javascript. 
\end{itemize}

\begin{figure}
\epsscale{.80}
\plotone{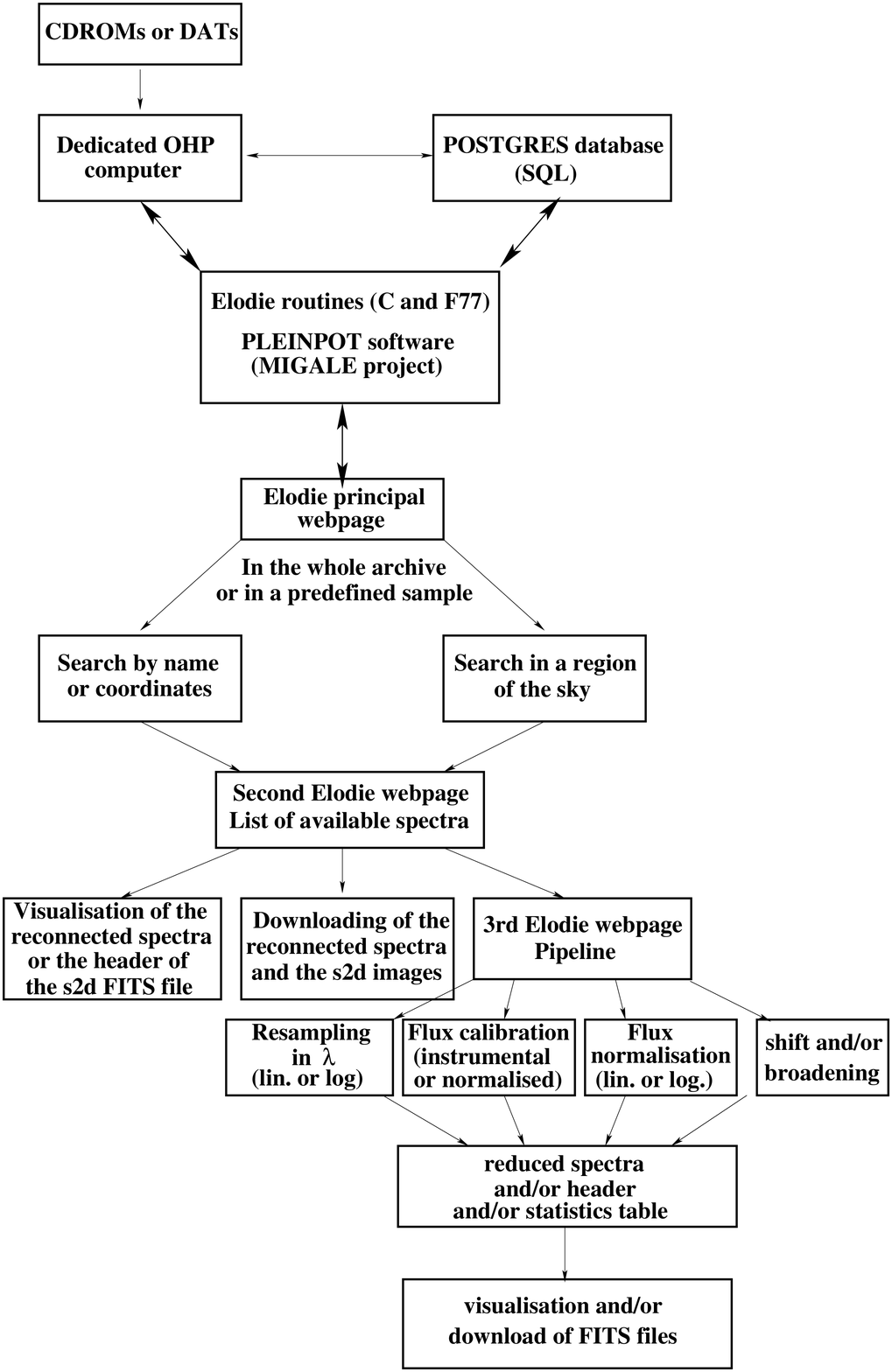}
\caption{Schematic organisation of the ELODIE archive.\label{organi}}
\end{figure}

In the following section we describe the user interface of the archive, the data products that are accessible through the webpages and the available data processing functions. The general organisation of the archive as well as the on-line facilities are summarized in the diagram shown in Fig.~\ref{organi}.

\section{User interface, data products and data processing}\label{sec:inter}

\subsection{The user interface}
The principal webpage of the archive is designed in a similar way as the SIMBAD principal webpage in order to make its use easier for the customers of the SIMBAD database.
Through this webpage, the archive can be interrogated using either (1) object naming conventions or (2) positions. 
In the case of positions, the search is carried in a region around the coordinates given by the user. If the request is made by name, the user can choose as well to search in a region around the object.

In preparing the archive we have strived to label as much science data as possible with a meaningful object name. In doing so we have relied first on the name and catalog coordinates given by the observer and second on the telescope
coordinates. The results may not always be satisfactory due to unrecognized or unavoidable errors. These problems are planned to be solved in the forthcoming future (see Sect. \ref{sec:perspectives}).

The first webpage of the archive interface also allows the user to make a search in a predefined list of objects. At present only two lists are available: the whole list of the archived objects and a list of stars with measured metallicities (Cayrel de Strobel et al. 1997, 2001).   


The results of a search are then listed in the second dedicated webpage of the archive. The obtained list provides the designation of the objects satisfying the request, their J2000 coordinates, a link to the SIMBAD webpage corresponding to each object, the date of observation and the number of observations made during the night. This webpage allows as well to visualise the header of the FITS file corresponding to an observation. Finally, through this interface, the user can  visualise or download the reconnected spectrum, described in the next section, download the s2d file also described hereafter or access the pipeline interface (Sect. \ref{sec:pipeline}).\\
 \\
Downloading the data:\\

\noindent The http interface has been designed to be accessed both by a human reader using a standard browser and by programs such as {\it wget}, a classical web crawler, or astronomical packages, such as Pleinpot. The possibility
to access the service by a program is a fundamental requirement for the Virtual
Observatory, and the interface will evolve to incorporate the new protocols
and data-model being discussed for the VO.\\
A description of the different possibilities to download the ELODIE archive data can be found in the help pages of the website.


\subsection{Data products}\label{sec:dataproducts}
\begin{figure}
\epsscale{.80}
\plotone{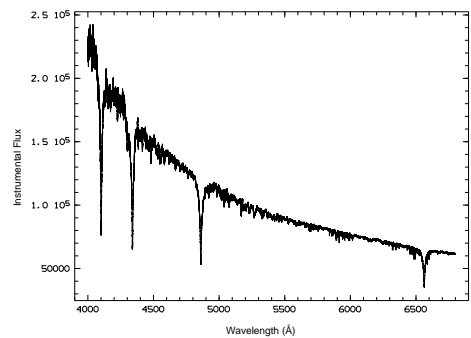}
\caption{The reconnected spectrum of Altair after correction
for the incomplete removal of the instrumental signature (free from the saw-tooth shape). The energy distribution is now relative to that of the tungsten lamp and thus appears ``too blue.''\label{spectrecor}}
\end{figure}

As it has been explained in the previous section, the second webpage allows the user to access three different products related to an observation:

1- The reconnected spectrum resampled in wavelength with 
a constant step of 0.05 \AA\, covering the range 4000 to 6800 \AA\ 
is given in ``instrumental'' flux. 
A procedure developed by Prugniel \& Soubiran (2001) has been
implemented in the archive software in order to correct the reconnected spectra for the saw-tooth shape due to the incomplete removal of the blaze function. Application of this procedure results in an  ``instrumental''  flux distribution which is relative to the tungsten lamp, and thus appears ``too blue'' (see example in Fig. \ref{spectrecor}).   
Details on the three FITS file extensions used are given on the webpage of the archive.\\
The spectra are at present in the topocentric reference frame, i.e. "as observed" (not corrected for the radial velocity of the star nor for the Earth's velocity around the Sun as projected in the line of sight to the star). These corrections will be made available in the next version of the archive software (see Sect. \ref{sec:perspectives}). 

2- The header of the s2d FITS image provides useful information such as the name of the observer, the telescope position, the S/N ratio, wavelength calibration, etc... 

\begin{figure}
\epsscale{.80}
\plotone{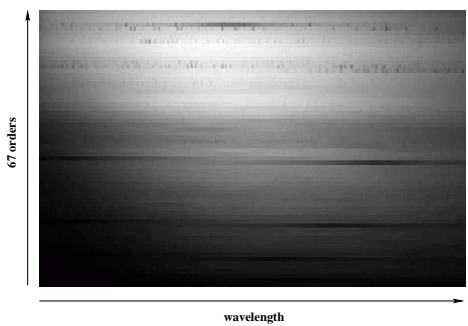}
\caption{s2d image for the observation of the star 
Altair.
The broad Hydrogen lines are clearly visible, particularly 
$H_\alpha$ close to the top. 
 The narrow lines in the upper part of the 
images are due to water vapor in the Earth's atmosphere.\label{s2d}}
\end{figure}

3- The s2d image is composed of 67 lines of 1024 pixels (see Fig. \ref{s2d}) holding the extracted spectra for all orders (covering the full spectral 
range), corrected for the blaze problem (i.e. reconnectable), 
where a wavelength is assigned to each pixel (the coefficients are given in the
FITS header). The s2d spectrum is corrected for the curvature of the echelle orders, but due to a bad correction for scattered light in the on-line reduction, a residual curvature can still remain. Such a correction is made in the wavelength resampled spectra (see above). The advantage of a spectrum in the s2d format is that the original (variable) pixel sampling is preserved across the entire wavelength interval, as no rebinning has taken place, but special software is needed to use this format. Details on the two FITS file extensions used are given on the webpage.\\


\subsection{The Pipeline interface}\label{sec:pipeline}
The webpage of the pipeline allows the user to generate the reconnected spectra on-the-fly. Four different functions are available for this purpose:

1- The resampling function which allows to resample the spectrum in wavelength with a constant step inside a wavelength interval chosen by the user. The available scales are linear and logarithmic.

2- The flux calibration function provides two choices at present: an ``instrumental flux'' calibration (defined in Sect. \ref{sec:dataproducts}) or a ``normalised continuum flux'' calibration.

3- The flux normalisation function allows to normalise the spectrum in a wavelength interval chosen by the user. The flux scale can be linear or logarithmic. 

4- The spectral shift and broadening function with a chosen velocity and dispersion. 


Data evaluation is made by visualising or downloading one of the three different outputs: the "statistics" table (number of pixels and lines, mean flux and the RMS), the "header" list and/or the "spectrum". 

The user has the possibility to choose the type of visualisation of the output among different options: a hypertext page, an HTML page and a download to a FITS file, a download to a gzipped FITS file, a download to xview and a download to a FITS viewer.


\section{Plans for future development}\label{sec:perspectives}

We plan to proceed to the development of phase 2 of the archive software in the summer of 2004, aiming principally to provide additional options for selecting and retrieving the data. New processing functions may also become available during this phase (such as a flux calibration process) and links with other projects will be developed.  \\
We list hereafter the main options that are planned to be improved  during the phase 2 development of the archive:

1- Searching for data by observational characteristics : The observational parameters, generated when the observations are obtained at the telescope and included in the FITS header (as the S/N ratio, the date of observation, the name of the observer etc...) are already stored in a table of the database. An advanced selection menu to put constraints on these parameters will be developed.

2- Searching for data by stellar characteristics: A "server of fundamental parameters of stars" is planned to be installed in 2004 at Bordeaux. It will provide the effective temperature, surface gravity and metallicity of stars (extension of the Cayrel de Strobel et al. (2001) catalogue) using the TGMET code of Katz et al. (1998). In addition, it will also provide estimates of the absolute magnitude, the spectral classification, the magnitude and the distance.\\
The latter parameters are available at present from the SIMBAD database and we
are currently studying improvements of cross-access with SIMBAD. Possibly, in the future, SIMBAD will give direct access to the entries in the ELODIE archive. Thanks to the Bordeaux server, the search for data in the ELODIE archive according to physical characteristics criteria will be possible.

3- Searching data using predefined lists of samples: The list of pre-defined samples will also be extended. 

4- Searching data from a list of objects provided by the user: In order to facilitate the use of the archive, we intend to provide the possibility for the user to upload his own list of objects in order to search for eventual spectra present in the archive.

5- Downloading a whole sample: We intend to provide a script allowing the user to download a list of spectra. This option is preferable rather than the use of the web interface, though this latter option is still a possibility.\\
 
In the meantime, the content of the ELODIE archive is continuously updated and efforts are undertaken to improve the quality assessment of the material and the quality of the cross-identifications.\\

The ELODIE spectrograph will be replaced in late 2005 by a new instrument, SOPHIE, with improved capabilities in terms of stability, limiting magnitude and accuracy of the radial velocities (an improvement of a factor of 2.5 on the magnitude and of 2.5 to 3 on the radial velocities). We plan to extend the present archive to host the spectra from SOPHIE.

\acknowledgments
We would like to thank the members od the Elodie archive Science Committee, A. L\`ebre, F. Ochsenbein, M. Haywood and J-P. Sivan for their help. \\
We are grateful to the Observatoire Astronomique Marseille-Provence (OAMP) and to Program (PNG) for financial support.\\
JM is thankful to A. Eckart and Univ. of Cologne for their support.



Facilities: \facility{OHP 1.93m telescope (ELODIE)}.



\appendix

\clearpage




\clearpage
\begin{table}[htbp]
\begin{center}
\begin{tabular}{rc}
\tableline\tableline
Total number of spectra &   \\
for the period 1994-2001 & $\sim 18000$ \\
\tableline
Number of public spectra & $\sim 9000$  \\
\tableline
Number of spectra &  \\
with $10 \leq S/N \leq 200$& $\sim 6000$ \\
\tableline
Number of spectra &   \\
with $S/N \geq 100$& $\sim 3000$ \\
\tableline
Spectral range & 3895\AA\ - 6815\AA\ \\
\tableline
Spectral resolution R & $\sim 42000$\\
\tableline\tableline
\end{tabular}
\end{center}
\caption{Characteristics of the archive. The S/N ratio is measured at 5500$\AA$.}
\label{tab:car}
\end{table}


\end{document}